\documentstyle[12pt]{article}
\newcommand{\rf}[1]{(\ref{#1})}

\newcommand{\ep}{\varepsilon}

\newcommand{\de}{\delta}
\newcommand{\Gm}{G_{\mu}}
\newcommand{\dm}{\triangle \mu}

\newcommand{\ra}{\right\rangle}
\newcommand{\la}{\left\langle}

\newcommand{\cO}{{\cal O}}

\newcommand{\noi}{\noindent}

\def\void{}
\def\labelmark{}

\newenvironment{formula}[1]{\def\labelname{#1}
\ifx\void\labelname\def\junk{\begin{displaymath}}
\else\def\junk{\begin{equation}\label{\labelname}}\fi\junk}%
{\ifx\void\labelname\def\junk{\end{displaymath}}
\else\def\junk{\end{equation}}\fi\junk\labelmark\def\labelname{}}

\newenvironment{formulas}[1]{\def\labelname{#1}
\ifx\void\labelname\def\junk{\begin{displaymath}\begin{array}{lll}}
\else\def\junk{\begin{equation}\label{\labelname}\left.
\begin{array}{lll}}\fi\junk}%
{\ifx\void\labelname\def\junk{\end{array}\end{displaymath}}
\else\def\junk{\end{array}\right.\end{equation}}
\fi\junk\labelmark\def\labelname{}\def\junk{}
}

\newcommand{\beq}{\begin{formula}}
\newcommand{\eeq}{\end{formula}}
\newcommand{\beqv}{\begin{formula}{}}
\newcommand{\bea}{\begin{formulas}}
\newcommand{\eea}{\end{formulas}}

\begin{document}
\topmargin 0pt
\oddsidemargin 5mm
\headsep 0pt
\topskip 9mm

\begin{center}
{\large UNIVERSALITY IN THE CRITICAL BEHAVIOR OF 
THE CORRELATION FUNCTIONS IN $2D$ SIMPLICIAL GRAVITY}\\
 
\vspace{1cm}
{\bf J. Jurkiewicz}\footnote{Permanent address: Institute of Physics,
Jagellonian University, ul. Reymonta 4, PL-30-059, Krak\'{o}w 16, Poland.
Partly supported by the KBN grant 2PO3B 196 02.}\\
\vspace{1cm}
Laboratoire de Physique Th\'{e}orique et Hautes Energies\footnote{
Laboratoire associ\'{e} au Centre National de la Recherche Scientifique
-- URA D0063}\\
Universit\'{e} de Paris XI, b\^{a}timent 211, 91405 Orsay Cedex,
France
\end{center} 
\vspace{3cm}
\noi
{\bf Abstract}

The analogue of the loop-loop correlation function in 2d gravity for the
planar connected $\phi^3$ diagrams is calculated. It is shown that although the
discretized formulas are different the scaling limit is the same as 
for the loop-loop correlation function. The derivation may serve as an
alternative definition of the volume-volume correlator of Euclidean
quantum gravity in $2d$.
\vfill
\noi
LPTHE Orsay 96-13\\
February 1996

\newpage

\section{Introduction}

The concept of the transfer matrix in the pure 2d gravity was introduced in 
\cite{wat1}. The authors considered the "time" evolution of a given (closed)
loop on a triangulated random lattice as a deformation of this loop by
one step in the "forward" direction. The deformation consisted of removing 
the triangles attached to the links of the loop and could be interpreted
as an evolution by the unit length in the discrete geodesic distance.

The formulation of the problem made extensive use of the disk amplitude
$F(x,g)$, which was obtained earlier in \cite{zuber} in the large-$N$
$\phi^3$ matrix model. $F(x,g)$ is a generating function of
Green's functions of the model:
\beq{1.1}
F(x,g) = 1 + \sum_{k=1}^{\infty} x^k G_k(g),
\eeq
where
\beq{1.2}
G_k(g) = \la Tr \phi^k \ra
\eeq
and the average is taken in the large-$N$ limit of the $\phi^3$ matrix
model.

Green's functions defined above in the language of the planar $\phi^3$
graphs contain both connected and
 disconnected diagrams. For loops drawn on a
dual, triangulated surface this means the existence of pathologies, like
two-fold links (corresponding to the disconnected bare two-point functions).
These pathologies are expected to be irrelevant in the scaling limit
of the theory.

In this note we show that this is indeed the case. We propose an alternative
derivation of the transfer matrix and loop-loop correlators based on the
connected $\phi^3$ diagrams. In this formulation all pathologies mentioned 
above are explicitly excluded. We show that the scaling limit of the
two approaches is exactly the same.

\section{Connected disc amplitude}

The generating function of the connected Green's functions (with one marked
external line) for the $\phi^3$ theory
\beq{2.1}
\Psi(x,g) = 1 + \sum_{k=1}^{\infty} x^k \Psi(k,g)
\eeq
was also considered in the paper \cite{zuber}. In the planar limit this
function satisfies the quadratic equation
\beq{fundeq}
g\Psi^2 - (x+g)\Psi + \frac{(s+1)(3s+1)}{8s}x + x^3 = 0,
\eeq
where the coupling constant $g$ is related to $s$ by
\beq{2.3}
g^2 = 8s(1-s^2).
\eeq
In this parametrization $g=0$ corresponds to $s=1$. The critical point
of the theory is given by
\bea{gc-sc}
g_c &=& \frac{1}{2~ 3^{3/4}},\\
s_c &=& \frac{1}{3^{1/2}}.
\eea

Equation \rf{fundeq} can easily be solved. One gets
\beq{solution}
\Psi(x,g)=\frac{1}{2}(\frac{x}{g}+1)+\frac{1}{2}(1-\frac{s x}{g})
\sqrt{1-\frac{4 g x}{s^2}}.
\eeq
For $g \to 0$
\beq{2.4}
\Psi(x,g=0)=1+x^2,
\eeq
where $x^2$ is the contribution from the bare propagator.
From \rf{solution} we find the critical value $x_c$
\beq{xc}
x_c = \frac{1}{2~ 3^{1/4}},
\eeq
which in the scaling region $g \to g_c$ is approached from {\em above}.

Note that the analytic structure of $\Psi(x,g)$ is quite different from that of 
$F(x,g)$. However the scaling behavior of the two functions turns out to be 
very similar. Introducing the parametrization 
\bea{params} 
g &=& g_c e^{-\ep^2 
t},\\ 
s &=& s_c (1 + \frac{2\ep\sqrt{t}}{\sqrt{3}}),\\ 
x &=& x_c (1 - \ep 
\zeta), 
\eea 
we get the formula very similar to that obtained for $F(x,g)$ in the paper 
\cite{wat1}:  
\beq{cr} 
\Psi = \frac{1+\sqrt{3}}{2}(1-\frac{3-\sqrt{3}}{2}\ep\zeta) 
+\frac{1}{4}f(\zeta,\tau)\ep^{3/2} + \cO(\ep^2).
\eeq
In \rf{cr}
\bea{tu}
f(\zeta,\tau) &=& (2~\zeta-\sqrt{\tau})\sqrt{\zeta+\sqrt{\tau}}\\
\sqrt{\tau} &=& \frac{4}{\sqrt{3}} \sqrt{t}.
\eea

The only difference is in the value of the numerical coefficients in
front of the $\ep$ and $\ep^{3/2}$ terms and in the definition of
$\tau$.

\section{The transfer matrix by the slicing method}

Since the connected $\phi^3$ diagrams rather then the triangulated 
surfaces will be the subject of this note, we have to modify the concept of 
the {\em slicing} to get the deformation laws. Let us consider the planar 
diagram with $L$ external lines. One of them is a marked line.  The deformation 
we shall consider will consist of eliminating all external links, together with 
vertices from which they emerge. After such operation we get again a connected 
diagram, but in general with a different length $L'$. 
Following the method proposed
in \cite{wat3} we consider the generating function 
\beq{4.1}
\Gm^{(0)}(x,y;1)=\sum_{L,L'}x^L y^{L'}\Gm^{(0)}(L,L';1),
\eeq
where $\Gm^{(0)}(L,L';1)$ is the sum of all possible graphs connecting
loops separated by one deformation layer with lengths $L$ and $L'$.
In this form the external lines are not marked. Similarly like in \cite{wat3}
the graphs can be obtained as a combination of five types of graphs (a) - (e).
\\
\unitlength=1.00mm
\special{em:linewidth 0.4pt}
\linethickness{0.4pt}
\begin{picture}(127.33,35.00)
\put(20.00,25.00){\circle*{2.00}}
\put(40.00,25.00){\circle*{2.00}}
\put(80.00,25.00){\circle*{2.00}}
\put(100.00,25.00){\circle*{2.00}}
\put(120.00,25.00){\circle*{2.00}}
\put(20.00,35.00){\line(0,-1){10.00}}
\put(20.00,25.00){\line(-5,-6){7.33}}
\put(20.00,25.00){\line(5,-6){7.33}}
\put(32.00,35.00){\line(4,-5){8.00}}
\put(40.00,25.00){\line(0,-1){9.00}}
\put(48.00,35.00){\line(-4,-5){8.00}}
\put(60.00,35.00){\line(0,-1){10.00}}
\put(60.00,25.00){\line(5,-6){7.33}}
\put(80.00,35.00){\line(0,-1){10.00}}
\put(80.00,25.00){\line(-5,-6){7.33}}
\put(100.00,35.00){\line(0,-1){10.00}}
\put(100.00,25.00){\line(-5,-6){7.33}}
\put(120.00,35.00){\line(0,-1){10.00}}
\put(120.00,25.00){\line(5,-6){7.33}}
\put(55.00,18.00){\circle{6.32}}
\put(85.00,18.00){\circle{6.32}}
\put(110.00,19.00){\oval(10.00,8.00)[]}
\put(60.00,25.00){\circle*{2.00}}
\put(60.00,25.00){\line(-1,-1){4.00}}
\put(80.00,25.00){\line(1,-1){4.00}}
\put(100.00,25.00){\line(1,-1){5.00}}
\put(120.00,25.00){\line(-1,-1){5.00}}
\put(20.00,5.00){\makebox(0,0)[cc]{(a)}}
\put(40.00,5.00){\makebox(0,0)[cc]{(b)}}
\put(60.00,5.00){\makebox(0,0)[cc]{(c)}}
\put(80.00,5.00){\makebox(0,0)[cc]{(d)}}
\put(111.00,5.00){\makebox(0,0)[cc]{(e)}}
\end{picture}

Their contributions are:
\bea{gr}
{\rm (a)} &=& gx^2 y,\\
{\rm (b)} &=& gx y^2,\\
{\rm (c)} &=& {\rm (d)} = g x y\frac{\Psi(x,g)-\Psi(x,g=0)}{x},\\
{\rm (e)} &=& g^2 x^2 y^2 
\frac{\Psi(x,g)-1-x\frac{\partial \Psi(x,g)}{\partial x}|_{x=0}}{x^2}.
\eea

Subtraction in (c) is necessary to exclude the contribution from the bare 
propagator. In (e) we have to exclude from $\Psi$ diagrams with
zero and one line.

Summation of these contributions gives
\bea{contr}
\Gm^{(0)}(x,y;1)&=&\sum_{n=1}^{\infty}\frac{1}{n}
({\rm (a)}+{\rm ((b)}+{\rm (c)}+{\rm (d)}+{\rm (e)})^n\\
&=& -\log(1-{\rm (a)}-{\rm ((b)}-{\rm (c)}-{\rm (d)}-{\rm (e)}).
\eea
where as in \cite{wat3} the factor $1/n$ results from the cyclic symmetry.

The transfer matrix at a unit distance is closely related:
\beq{transfer}
\Gm(x,y;1)=y\frac{\partial}{\partial y}\Gm^{(0)}(x,y;1).
\eeq
The derivative has the effect of marking one of the original external lines
and providing a correct contribution from this line (compare  ref. \cite{wat1}).

In the scaling limit the parameters behave as \ref{params}. For $y$ we
take
\beq{yc}
y = \frac{1}{x_c}(1-\ep \zeta').
\eeq
In the small $\ep$ limit we get
\beq{split}
\Gm(x,y;1)=\frac{1}{\ep}
\frac{1}{\zeta+\zeta'-\ep^{1/2}\alpha f(\zeta,\tau)}.
\eeq
Again the only difference is in the finite coefficient $\alpha$
\beq{alpha}
\alpha=\frac{2\sqrt{3}-1}{11}.
\eeq
This form of $\Gm(x,y;1)$ leads to the continuum differential equation
for the transfer matrix $\Gm(x,y;r)$ in the scaling limit. We shall not discuss it here, but
rather we rederive this equation using the {\em peeling} method.

\section{Equation for $\Gm(x,y;r)$ - the peeling method}

The idea of peeling was introduced in \cite{wat2}. In our context it corresponds
to the following deformation of the connected diagram: we start with arbitrary
external link and cut away this link together with the corresponding vertex.
This operation can be represented by the diagrams presented
before:
\beq{sli}
2~{\rm (a)} + {\rm (b)} +{\rm (c)} + {\rm (d)}.
\eeq
Notice the factor 2 in diagram (a), resulting from two external lines
and the absence of diagram (e), which require two consecutive cuts.  Notice 
also that a contribution of the
diagram (a) is completely cancelled by subtractions of the bare 
propagator contribution in (c) and (d).

The peeling operation can be repeated iteratively around the connected diagram.
Let us consider the function $\Gm(L,L';r)$, where $L'$ is the initial
number of external lines (at a distance $r$). The peeling changes $\Gm$
\beq{change}
\Gm(L,L';r) \to g \Gm(L+1,L';r) + 2g\sum_{L''=1}^{L+1}\Psi(L'',g)
\Gm(L-L''+1,L';r).
\eeq
This can be put in a form of a differential equation
\bea{no1}
\frac{1}{L}\frac{\partial}{\partial r}\Gm(L,L';r) &=& g \Gm(L+1,L';r)\\
 &+& 2g\sum_{L''=1}^{L+1}\Psi(L'',g)
\Gm(L-L''+1,L';r) - \Gm(L,L';r).
\eea
Multiplying by $L x^L y^{L'}$ and performing summation over $L$ and $L'$
we get
\beq{no2}
\frac{\partial}{\partial r} \Gm(x,y;r)=x\frac{\partial}{\partial x}
\left( (\frac{g}{x}-1+\frac{2g}{x}(\Psi(x,g)-1))\Gm(x,y;r)\right).
\eeq
Recalling the form of $\Psi(x,g)$ \rf{solution} we get
\beq{no3}
\frac{\partial}{\partial r} \Gm(x,y;r)=x\frac{\partial}{\partial x}
\left(\frac{g}{x}f_{\mu}(x,g)\Gm(x,y;r)\right),
\eeq
where
\beq{fmu}
f_{\mu}(x,g)=(1-\frac{s x}{g})\sqrt{1-\frac{4g x}{s^2}}.
\eeq
The explicit solution of this equation is
\beq{sol}
\Gm(x,y;r)=\frac{f_{\mu}(\hat{x})}{f_{\mu}(x)}\frac{1}{1-\hat{x}y},
\eeq
expressed in terms of the solution
$\hat{x}(x,r)$ of the characteristic equation. We have
\bea{char}
r &=& \int_x^{\hat{x}(x,r)}\frac{dx'}{g~f_{\mu}(x')}
=- \frac{1}{s\sqrt{1-\frac{4g^2}{s^3}}}
\log\frac{t(x')-\sqrt{1-\frac{4g^2}{s^3}}}
{t(x')+\sqrt{1-\frac{4g^2}{s^3}}}|_x^{\hat{x}(x,r)}\\
t(x')&=&\sqrt{1-\frac{4gx}{s^2}}.
\eea

Formula \rf{char} can easily be inverted. Here let us introduce notations
\bea{nota}
t &=& \sqrt{1-\frac{4gx}{s^2}},\\
\hat{t} &=& \sqrt{1-\frac{4g\hat{x}}{s^2}},\\
\de_0 &=& \frac{s}{2}\sqrt{1-\frac{4g^2}{s^3}}.
\eea
We have
\beq{important}
\hat{t}=\frac{2\de_0}{s}\frac{\coth(\de_0 r) +\frac{2\de_0}{ts}}
{1 + \frac{2\de_0}{ts}\coth(\de_0 r)}.
\eeq
This formula will be very important to study the scaling behavior of the 
transfer matrix. To agree with conventions used in \cite{wat3} we have 
\bea{scale}
g &=& g_c e^{-\dm},\\
s &=& s_c(1+\frac{2}{\sqrt{3}}\sqrt{\dm}),\\
\de_0 &=& \sqrt{6}g_c (\dm)^{1/4}.
\eea
In the scaling region obviously $\de_0 \to 0$.
Provided $x$ is not in the scaling region and $r$ is not to small we get
\beq{imp1}
\hat{t}=\frac{2\de_0}{s}\coth(\de_0 r) + \cO (\de_0^2).
\eeq

In the results presented
above notice that $\de_0$ has {\em exactly} the same value
as in \cite{wat3} and although the formulas are different the universal
large-$r$ behavior of $\Gm(L,L';r)$ is the same. We show it explicitly 
calculating the analogue of the two-point function $\Gm(r)$ which measures
the number of links a distance $r$ from a loop $L \to 0$. This function
can be expressed in terms of the two-loop function $\Gm(L,L';r)$ and the
one-loop function $\Psi(L',g)$.
We have
\bea{gog}
\Gm(r) &=& \sum_{L'=1}^{\infty}\Gm(L=0,L')L'\Psi(L')\\
&=& \oint_{C_y}\frac{dy}{2\pi{\rm i}y}\Gm(0,\frac{1}{y};r)y\frac{\partial}
{\partial y}\Psi(y)\\
&=&f_{\mu}(\hat{x})\hat{x}\frac{\partial}{\partial\hat{x}}\Psi(\hat{x})|_{x=0}.
\eea
In the derivation we made use of $f_{\mu}(0)=1$. As was shown above in the
scaling limit $\hat{t}\to 0$, so 
\beq{sc1}
\hat{x}\frac{\partial}{\partial\hat{x}}\Psi(\hat{x})|_{x=0} \to
2~\sqrt{3} + \cO(\de_0^2)
\eeq
is dominated by the nonuniversal part of $\Psi$.
Expressing $f_{\mu}(\hat{x})$ in terms of $\hat{t}$ we have
\beq{sc2}
f_{\mu}(\hat{x})=\frac{s^3}{4g^2}\hat{t}(\hat{t}^2 - \frac{2\de_0^2}{s}),
\eeq
and in the scaling limit we get
\beq{final}
\Gm(r)=36~\de_0^3\frac{\cosh (\de_0 r)}{\sinh^3(\de_0 r)}(1 + \cO(\de_0)).
\eeq
Up to the numerical constant in front this is precisely the form obtained
in \cite{wat3}.

\section{Discussion}

The calculation presented above is a nice demonstration of the universal
character of the scaling limit in 2d simplicial gravity. Results were
to large degree expected to agree with results of \cite{wat1,wat2,wat3},
however the emergence of the same asymptotics from apparently different
discretized forms is a strong confirmation of these expectations.

The calculation is in many points even simpler due to the simpler analytic
structure of the connected disc amplitude $\Psi(x,g)$. The scaling part
of this function has almost immediately the right form. 

The description in terms of $\phi^3$ graphs was used in the whole paper.
It is obvious that the same could be achieved using triangles as in
\cite{wat1,wat2,wat3}. The contribution both in the slicing and in the peeling
methods are however 
different than in these references, to take into the account the
fact that the disc's boundary has no singularities.

\vspace{1cm}
\noi
{\bf Acknowledgements}

The author would like to thank LPTHE for the hospitality during his stay
there and A. Krzywicki for many stimulating discussions.

\end{document}